\documentclass[useAMS,usenatbib]{mnras}

\usepackage{tabularx}
\usepackage{multirow}
\usepackage{url}
\usepackage{amsmath}	
\usepackage{amssymb}	
\usepackage{aas_macros}
\usepackage{graphicx}

\usepackage{times,txfonts}
\usepackage{bm}

\usepackage[T1]{fontenc}
\usepackage{ae,aecompl}
\newcommand\lb{\langle}
\newcommand\rb{\rangle}

\usepackage{dcolumn}
\usepackage{hyperref}
\usepackage{xcolor}
\usepackage{soul}

\newcommand{\ramses}{{\sc ramses}}          
\newcommand{\vapor}{{\sc vapor}}          
\newcommand{\Fig}[1]{Fig.~\ref{fig:#1}}    
\newcommand{\Figure}[1]{Figure~\ref{fig:#1}}    

\title[Young magnetised discs]{The effects of large scale magnetic fields around young protostars and their disks}

\author[M. Kuffmeier and F. Nauman]{Michael Kuffmeier$^1$\thanks{E-mail: kueffmeier@nbi.ku.dk}
	Farrukh Nauman$^{2}$\thanks{E-mail:	nauman@nbi.ku.dk} \\
	$^{1}$Centre for Star and Planet Formation, Niels Bohr Institute and Natural History Museum of Denmark, University of Copenhagen, \O ster Voldgade 5-7,DK-1350 Copenhagen K, Denmark. \\
	$^{2}$Niels Bohr International Academy, The Niels Bohr Institute, Blegdamsvej 17, DK-2100, Copenhagen \O, Denmark.}

\pubyear{2017}

\begin{document}
	
	\date{\today}
	
	\pagerange{\pageref{firstpage}--\pageref{lastpage}}
	
	\maketitle
	
	\label{firstpage}
	
	\begin{abstract}
		We study the evolution of magnetic fields in accretion flows around six different sink particles in a global star formation simulation. Using the adaptive mesh refinement capabilities of \ramses, 
		the vicinity of the sinks is resolved with a resolution down to $2$ AU. The statistical properties of the magnetic field are tracked as a function of radius, height and time around each sink particle. All six systems are strongly magnetised with plasma beta being unity or below and we know that at least three of the sinks host an accretion disc. One of the discs is studied at a higher resolution of $0.06$ AU and we report its magnetic properties. We find that the angular momentum transport is dominated by large scale radial-azimuthal Maxwell stress. Furthermore, contrary to the weakly magnetised case studied in shearing box simulations, the large scale azimuthal field does not show smooth cycle periods but instead forms a banded structure. We conclude by speculating what this result might mean for observations and whether it will hold with highly resolved simulations including turbulence or non-ideal MHD effects.
	\end{abstract}
	

\date{\today}

\maketitle

\section{Introduction}  
Discs around young stars occur ubiquitously \citep[e.g.][]{Brinch2007,Yen2014,Tobin2015,Ansdell2016}. 
They form due to angular momentum conservation during the collapse of a prestellar core \citep[e.g.][]{Tscharnuter1975,Kamiya1977}. 
However, owing to flux freezing, 
an initially magnetised cloud tends to accumulate very strong magnetic fields that can potentially inhibit the formation of discs \citep{Allen2003,Galli2006,Hennebelle2008}.
This problem is commonly referred to as the `magnetic braking catastrophe' \citep{Mellon2008}. 
Previous works have shown different mechanisms to circumvent this problem, 
such as turbulence \citep{Seifried2012,Seifried2013,Joos2013,Li+2014,kuffmeier}, 
misalignment of the magnetic field with the angular momentum vector \citep{Hennebelle_Ciardi2009,Joos2012,Krumholz2013} or non-ideal magnetohydrodynamics (MHD) effects \citep{Machida2011,Tomida2015,Tsukamoto2015a,Tsukamoto2015b,Masson2016,Hennebelle2016,Wurster2016}.

The question of what transports the angular momentum thus allowing for mass to accrete onto the star is an outstanding problem of crucial importance for star and planet formation. 
The source of transport could be hydrodynamic leading to radial azimuthal stresses \citep{lovelace1999, klahr2003, marcus2015, shiavila2017}. 
Other possibilities include magnetohydrodynamic transport mechanisms such as the one due to weak magnetic fields (magnetorotational instability (MRI): \cite{1959velikhov}, \cite{1960PNAS...46..253C}, \cite{1991ApJ...376..214B}) 
and/or disc winds \citep{blandford1982,1993ApJ...410..218W}. 
A third possibility of transport through intermediate scale magnetic fields that are comparable or stronger than the thermal pressure \citep{1969Natur.223..690L,fieldrogers1993,pariev2003} has been explored in the context of active galactic nuclei. 

Recently, \citet{Kuffmeier2017B,kuffmeier} (hereafter referred to as K17a and K17b) carried out molecular cloud simulations with several sink particles. These numerical simulations solve the equations of ideal MHD starting from the molecular cloud scale and, using adaptive mesh refinement, manage to resolve scales as small as $2$ AU for most of the sink particles, and $0.06$ AU for one model. The circumstellar accretion discs are strongly magnetised. Strongly magnetised discs have been studied before in local simulations \citep{johansenlevin2008,salvesen2016} and global simulations \citep{gaburov2012,fragile2017} but mostly in the context of X-ray binaries and active galactic nuclei. 

This paper focuses on the magnetic field properties in the systems studied in more detail in K17a and is organised as follows. In section 2, we briefly describe the numerical setup. 
In section 3.1, we analyse some of the magnetic field properties around six systems we chose for zoom-ins with minimum resolution of $2$ AU. In section 3.2, we discuss in more detail one of the accretion discs that we zoomed into using a minimum cell size of $0.06$ AU.
We compare our work with local and global disc simulations in section 3.3. The conclusions are presented in section 4.

\section{Overview of the underlying simulations}
The analysis presented in this paper is based on simulations reported in K17a and K17b, which we refer the reader to for a more detailed description of the methods. 
The simulations are carried out with a modified version of the ideal MHD version of the adaptive mesh refinement (AMR) code \ramses\ \citep{Teyssier2002,Fromang2006}. We model the dynamics of a turbulent Giant Molecular Cloud (GMC) as a cubic box of ($40$ pc)$^3$ in volume with periodic boundary conditions. 
The average number density is 30 cm$^{-3}$ corresponding to about $10^5$ M$_{\rm \odot}$ of self-gravitating magnetised gas. 
The initial magnetic field strength is 3.5 $\mu$G and we use sink particles as sub-grid models for stars 
(please refer to \citet{Kuffmeier2016} and \citet{2017arXiv170901078H} for a description of the sink particle algorithm).
We drive the turbulence associated to a supernova explosion into the cloud, which results in a velocity dispersion of the cold dense gas that is in agreement with Larson's velocity law \citep{Larson1981}. Following the recipe of \citet{Franco_Cox_1986}, UV-induced heating \citep{Osterbrock_Ferland_2006} of the cold gas is quenched in the GMC. In lower density regions, where cooling becomes important, we apply a cooling function based on the work by \citet{Gnedin_Hollon_2012}. Using adaptive mesh refinement down to $16$ levels of $2$ with respect to the length of the box,  corresponding to a minimum cell size of $2^{-16} \times 40$ pc $\approx$ $126$ AU, we evolve the GMC for about 5 Myr. During this run, referred to as the parental run, several hundred sink particles form and evolve to different stellar masses. 

To gain further insight in to the formation process of the individual stars, K17a selected six sink particles for further analysis.
The six sinks are located in different environments of the GMC and by the end of the parental run, 
they have accreted about 1 to 2 $M_{\odot}$.
In the next step, we apply a higher resolution of $2$ AU around the sink of interest to follow the accretion process and the formation of the disc 
for about $50$ to $200$ kyr after star formation. 
Additionally, we apply a minimum cell size of $0.06$ AU for the disc of one of the sinks for an interval of about $1000$ years at the time of $\approx 50$ kyr 
after sink particle formation.
We point out that in all of these Zoom-in simulations, we still account for the full domain of the GMC (i.e. the entire box of ($40$ pc)$^3$ in volume).

\section{Magnetic field properties around forming stars}
In this section, we present the magnetic field properties during the formation phase of the six protostars selected for zoom-ins.
As described in detail in K17b, three of the selected objects host Keplerian discs that formed about $10$ kyr after sink formation. 
Two of these discs (sink b and sink d) have an approximately constant mass of ($M_{\rm disc} > 10^{-2}$M$_{\odot}$) and a disc radius of roughly $50$ to $100$ AU. 
The disc mass of the third object (sink f) drops from initially $\sim 10^{-2}$ M$_{\odot}$ to about $10^{-3}$ M$_{\odot}$ after $t=50$ kyr, 
while its disc radius is constantly about $20$ AU. The other three sinks do not host stable discs (sink c and sink e) or only host intermittent weak discs (sink a).
Considering that we -- opposite from previous global disc models \citep[e.g.][]{2015ApJ...801...84G,2017ApJ...845...75B,2017ApJ...836...46B} -- 
do not account for non-ideal MHD effects in our simulations, we possibly overestimate the magnetic field strength in the vicinity of the protostars.  
Despite this potential overestimation, a comparison of the ratio of kinetic to magnetic energy inside the discs shows
that the discs are rotationally dominated in particular in their inner parts (K17b and see also \citet{Nordlund2014}).
We present the properties and evolution of parameters such as plasma-$\beta$ and the spatiotemporal evolution of the toroidal component of the magnetic field $B_{\phi}$ in the six systems. To constrain the properties of a globally inherited magnetic field inside a young disc further, we investigate the magnetic properties of the disc around sink b in more detail in the high resolution run with a minimum cell size of $0.06$ AU.  

\subsection{Six zoom-in systems (2 AU)}
\begin{figure*}
	\centering
	\includegraphics[width=0.95\textwidth]{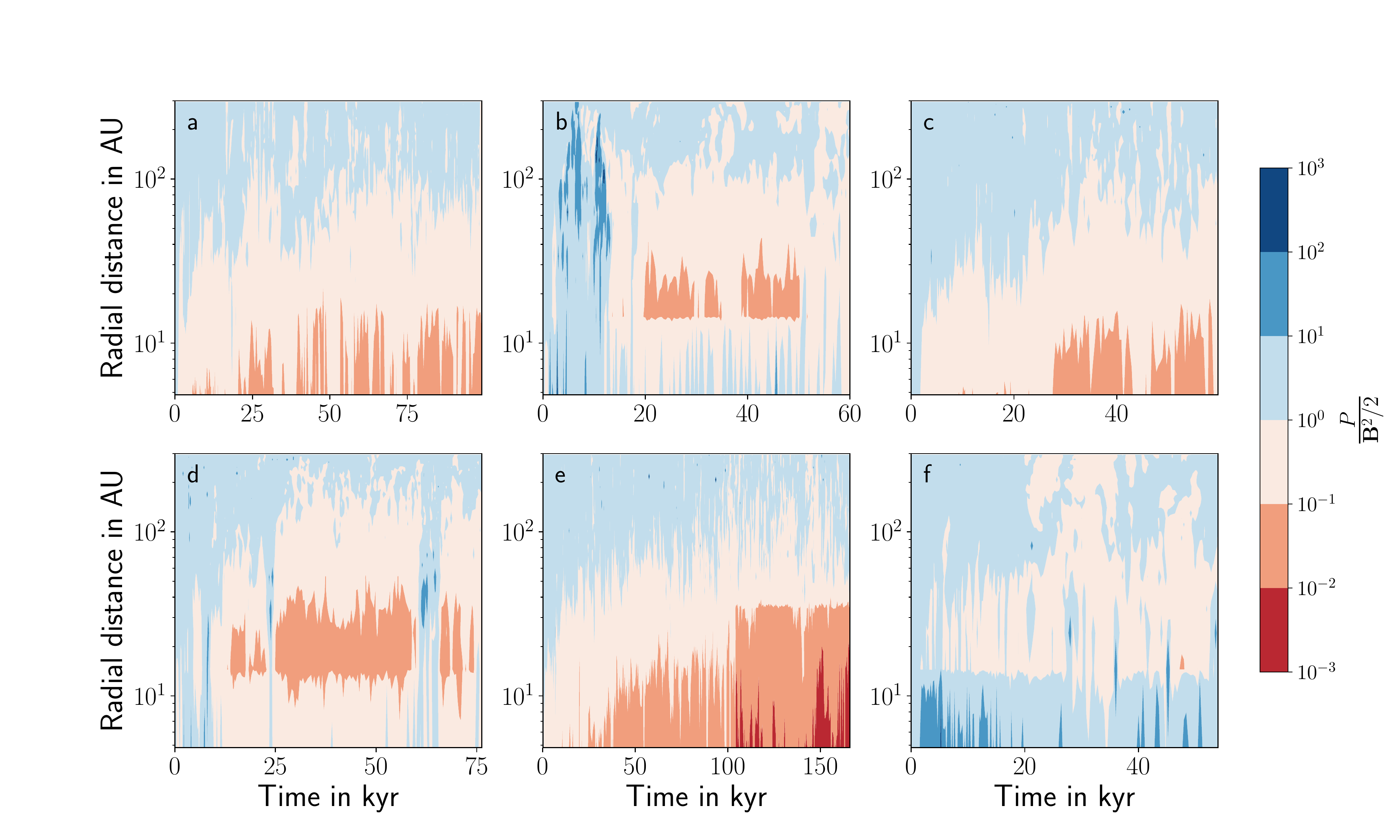}
	\caption{Contour plot of mass-weighted plasma beta, $\beta = 2P_{\rm th}/B^2$ and in the $t-r$ plane for six different runs. The models $(b)$, $(d)$ and $(e)$ develop a disc while the others do not. In all cases, the plasma beta is generally very low except for case $(e)$ where the plasma beta is considerably greater than unity at small radii.}
	\label{fig:beta}
\end{figure*}

In \Fig{beta} we show the ratio between thermal and magnetic pressure 
\begin{equation}
\beta = \frac{P_{\rm th}} {{B^2}/2}
\label{eq:beta}
\end{equation}
around the six sinks computed within cylindrical shells of radial width $\Delta r \approx 10$ AU 
and vertical extent $\Delta z = \pm 10$ AU. 
Beyond distances of $\sim 100$ AU, $\beta \gtrsim 1$ for all systems.
Comparing the six cases with each other, we find that the variations of $\beta$ at $R<100$AU are only minor.  
For the cases without clear signs of Keplerian discs (sink a, c and e), we find that $\beta$ is generally lowest close to the sink ranging typically in between $0.01$ and $0.1$ and increases with growing distance from the central sink to values $>1$ at $R\gtrsim 10$ AU. 
In contrast, $\beta$ is typically lowest at around $20$ AU from the central sink for the other three cases with a disc
with typical minimum values of around $0.1$ for sink b, $0.01$ to $0.1$ for sink d and $0.1$ to $1$ for sink f.
The disc around sink f is generally less magnetised than the other two discs, and in contrast to the other five cases $\beta$ is mostly $>1$ within the inner $\approx10$ AU from the central sink. 
$\beta$ within the inner $10$ AU of sink b is fluctuating around a mean value slightly lower than $1$, while $\beta$ is mostly $0.1$ -- $1$ within $10$ AU for sink d. 

\begin{figure*}
	\centering
	\includegraphics[width=0.95\textwidth]{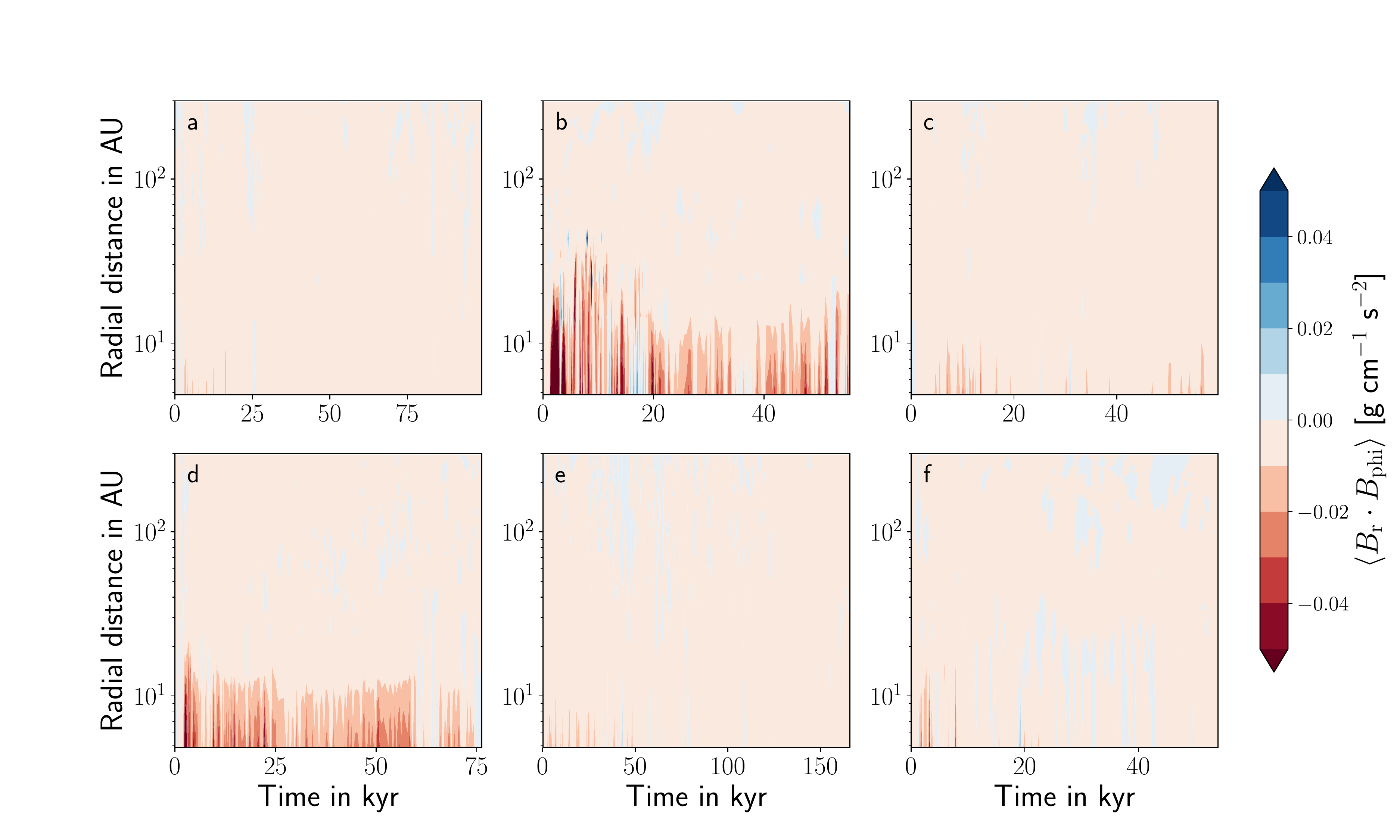}
	\caption{Contour plot of mass-weighted $\langle B_r \cdot B_\phi \rangle$ in $t-r$ plane for six different runs. The models $(b)$, $(d)$ and $(e)$ develop a disc while the others do not.}
	\label{fig:BrBphi}
\end{figure*}

\begin{figure*}
	\centering
	\includegraphics[width=0.95\textwidth]{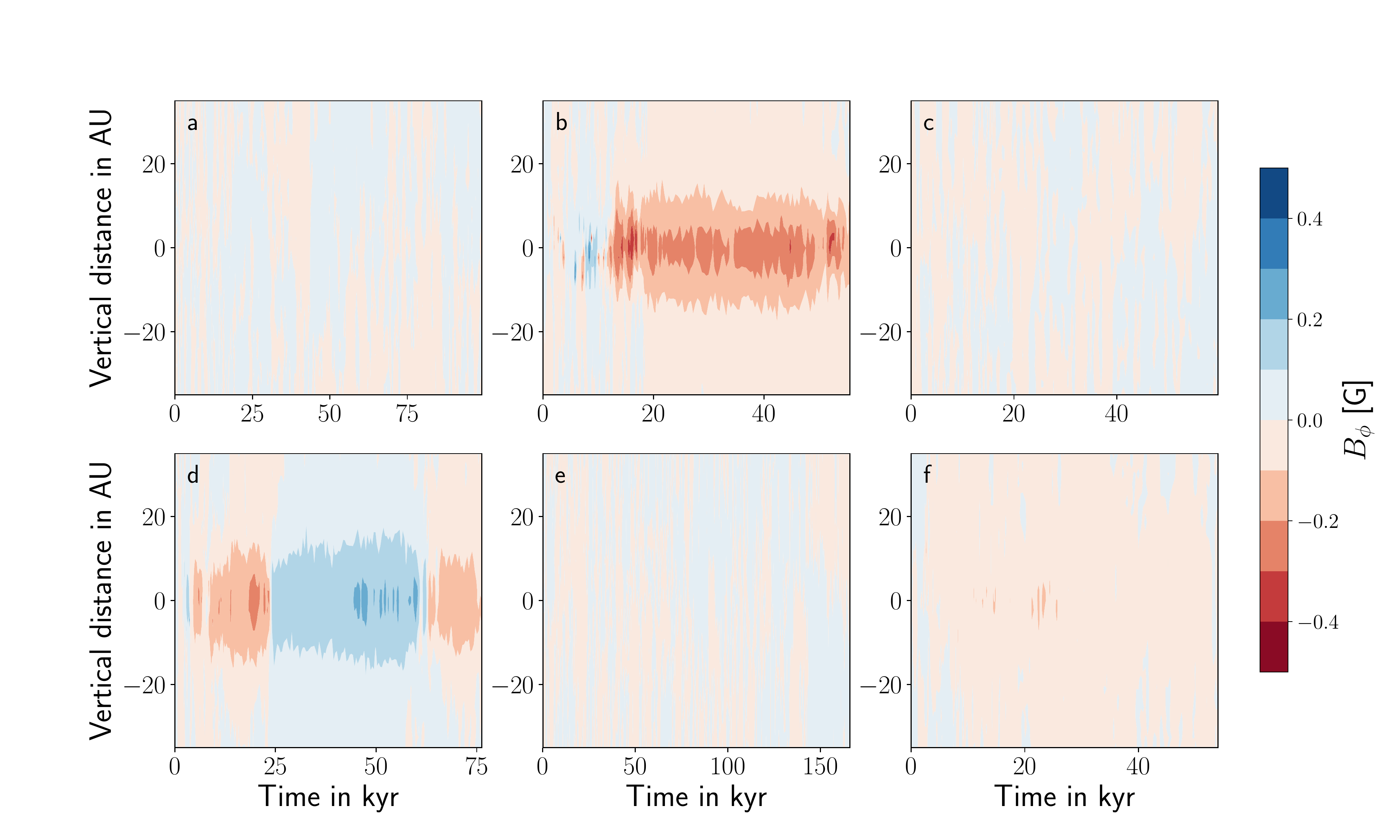}
	\caption{Contour plot of the mass-weighted azimuthal magnetic field in the $t-z$ plane for six different runs. The fields settle in a band like structure for the models $(b)$ and $(d)$ while showing a sign flip every few kyr. This is consistent with strong field local simulations by \protect\citet{salvesen2016} where a similar banded structure was seen for strong net vertical flux simulations. }
	\label{fig:bphi}
\end{figure*}

Especially at small radii, we expect larger $\beta$ values if radiative heating from the protostar was included.  
We find brief intervals lasting of the order of about $100$ -- $1000$ years, when $\beta$ suddenly increases from 
$\beta<1$ to $\beta>1$ at distances of about $20$ AU to $50$ AU.
These events occur most violently around sink d at $t\approx 24$ kyr and after $t\approx 60$ kyr, when $\beta$ rapidly increases from less than $0.1$ to more than $10$.
The rise in $\beta$ is induced by a rapid local increase in temperature, while the mean magnetic field $|\mathbf{B} |= \sqrt{B_x^2 + B_y^2 + B_z^2}$ remains approximately constant. 
Nevertheless, these events have direct consequence on the magnetic field structure as can be seen in \Fig{bphi}, where we show the mass-weighted average of the azimuthal magnetic field component at a radial distance around $22$ AU with $\Delta r \approx 4$ AU and $\Delta z \approx 2.3$ AU. 
The rapid increases in $\beta$ around sink d at $t\approx 24$ kyr and $t\approx 60$ kyr correlate with reversals of the magnetic field orientation.
The magnetic field reversal correlates with a sign flip of the radial B-field component, such that $\langle -B_r B_{\phi} \rangle > 0$ so that transport is predominately still outwards as seen in \Fig{BrBphi}.
$\langle B_r B_{\phi} \rangle < 0$ corresponds to outward transport of angular momentum in the disc (K17b).
In particular, we found that the radial-azimuthal Maxwell stress computed as 
\begin{multline}
F_r^B(R) = - \int_{-h/2}^{h/2} \mathrm{d}z \int_0^{2\pi} R\,\mathrm{d}\phi\, R \frac{B_{\phi}(R,\phi,z)B_r(R,\phi,z)}{4\pi}
\end{multline}
is the main contributor for outward transport, while the mechanical component (Reynolds stress) 
\begin{multline}
F_r^v(R) = \int_{-h/2}^{h/2} \mathrm{d}z \int_0^{2\pi} R\,\mathrm{d}\phi\, R\, \rho\, v_\phi(R,\phi,z)v_r(R,\phi,z) 
\end{multline}
is predominantly responsible for inward transport.
The gravitational component associated to transport in spiral arms is only of secondary importance.

Such banded structure and abrupt sign flips of the azimuthal magnetic field component are in agreement with results found by \citet{salvesen2016} in local shearing box simulations of highly magnetised discs, which they attributed to an artifact of periodic boundaries. In contrast, the magnetic fields in our simulations are anchored in the surrounding molecular cloud structure and are not periodic in the vertical direction. 
Sign flips have not been observed in recent global disc simulation with a weak magnetic field ($\beta \sim 10^3$: \cite{zhustone2017}), but these simulations were only evolved for a couple of thousand years as opposed to our simulations that have been evolved for close to $100,000$ years. 

Due to the coupling of magnetic fields and fluid, the two massive discs (sink b and sink d) also show the strongest azimuthal magnetic field close to the midplane, 
while the less massive disc around sink f has a lower toroidal field in agreement with the results of $\beta$. 
The remaining three cases without discs consistently do not show signs of a strong ordered toroidal magnetic field. 
Considering the evolution of the magnetic field in the cases with discs, we find that the amplitude of magnetic field $|\bm{B}|$ remains approximately steady. In other words, the magnetic energy does not decay when accounting for the global magnetic fields to which the toroidal fields in the discs are anchored. 

\subsection{Highly resolved disc (0.06 AU)}

\begin{figure}
	\centering
	\includegraphics[width=0.95\linewidth]{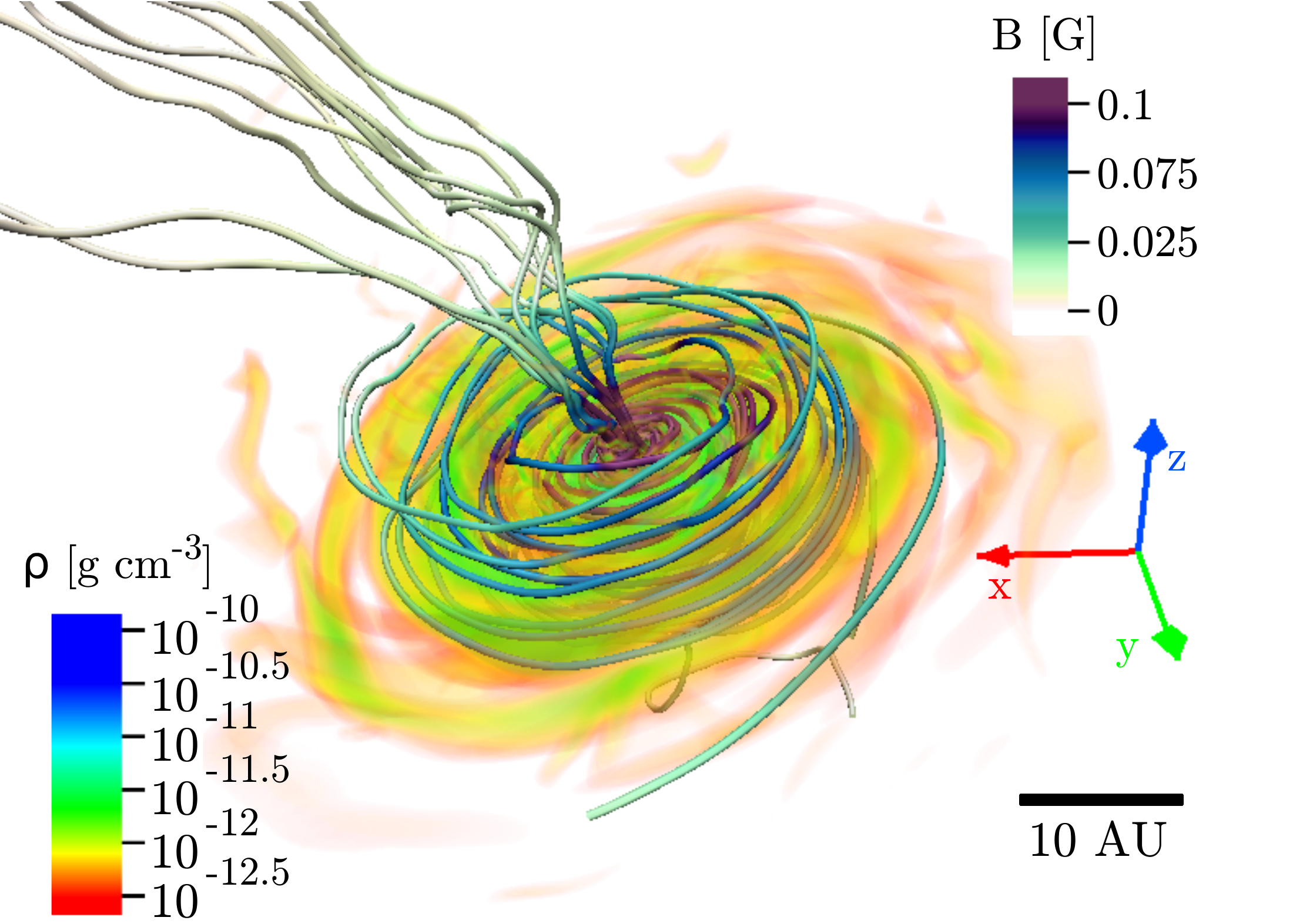}
	\caption{Visualisation of the disc and the corresponding magnetic field for the high resolution run of sink b after $t\approx 51$ kyr. We use the bias option in \vapor\ to only 
		visualise strong fields. The bias is set to 15, which means that $N\times 2^{15}$ seeds are computed 
		by the \vapor\ software of which the strongest $N$ are shown. We set the number of shown field lines to $N=10$. 
	}
	\label{fig:vapor_vis}
\end{figure}

To constrain the properties of the discs further, we apply additional refinement around sink b to reach a minimum cell size of $0.06$ AU. 
The additional refinement is applied gradually to let the model relax, and is applied at t = 50 kyr. The high resolution model is evolved for a $1000$ year period.  
Considering the sink mass of $M_*\approx 0.3$ M$_{\odot}$ at this stage, 1000 years correspond to more than 1150 orbital times at a fiducial radius of $0.6$ AU.
To give a better idea of the magnetic field structure inside the disc in three dimensions, we illustrate the disc and the corresponding toroidal magnetic field for one snapshot with the visualisation tool \vapor\ \citep{clyne2005prototype,clyne2007interactive}
in \Fig{vapor_vis}. 
Note that the opacity increases with increasing density and that densities less than $10^{-12.5}$ g cm$^{-3}$ are entirely transparent.
In this way, we can show both the magnetic field structure as well as the denser inner part of the disc. 
The global magnetic field lines are dragged in with the fluid and get wrapped up in a toroidal structure due to strong Keplerian rotation.
Moreover, the illustration shows that the magnetic field lines are anchored to the larger scales in vertical direction at small radii. 
We do not see a strong vertical outflow during the high zoom-ins, which is probably due to a significant contribution of vertical infall onto the system even at small radii at this point in time as discussed in K17a. 
The magnetic field strength is strongest in the vicinity of the star and weaker at larger distances. 

\begin{figure}
	\centering
	\includegraphics[width=0.475\textwidth]{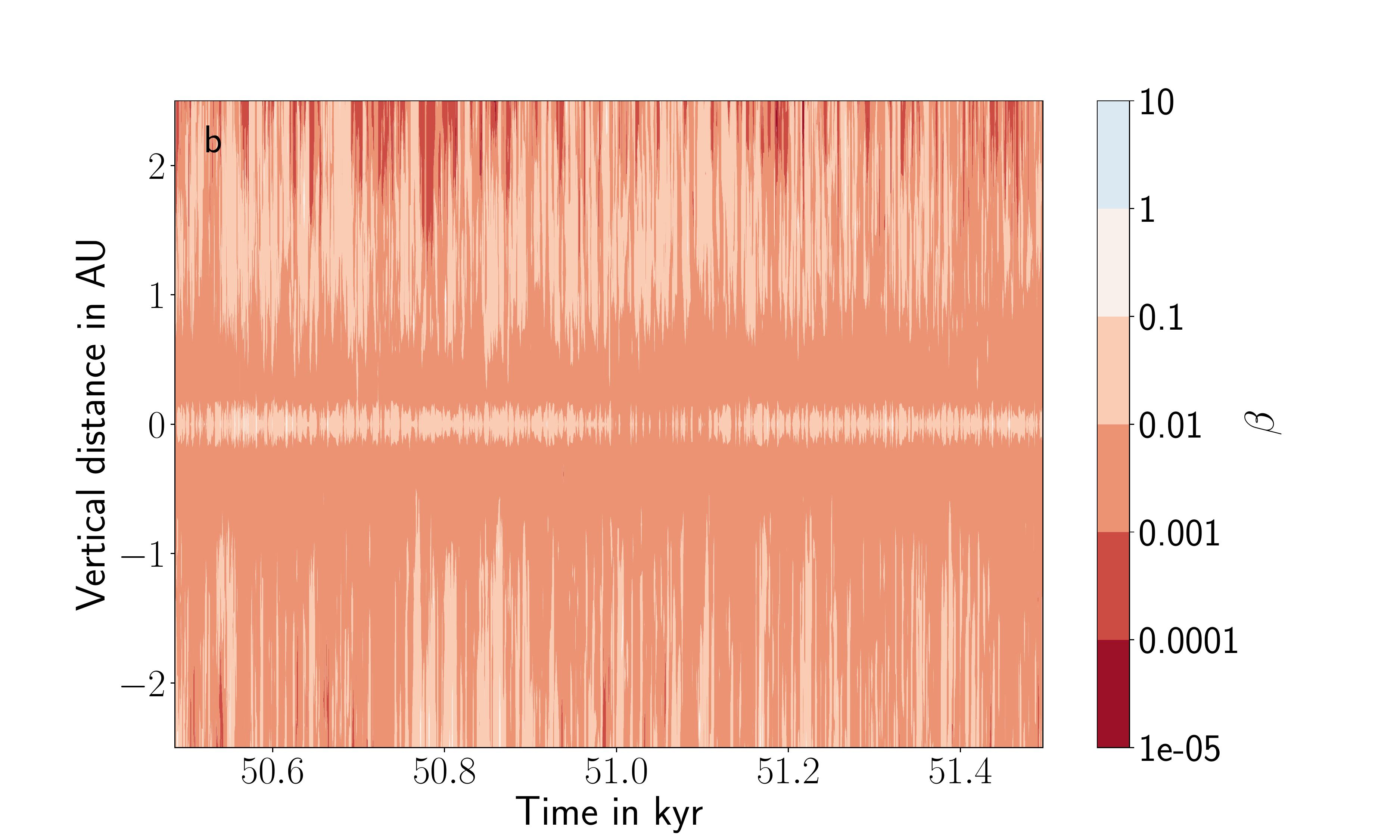}
	\caption{Contour plot of $\beta = 2p_{th}/B^2$ weighted by mass in the $t-z$ plane for the high resolution run.}
	\label{fig:betahr}
\end{figure}

\begin{figure}
	\centering
	\includegraphics[width=0.45\textwidth]{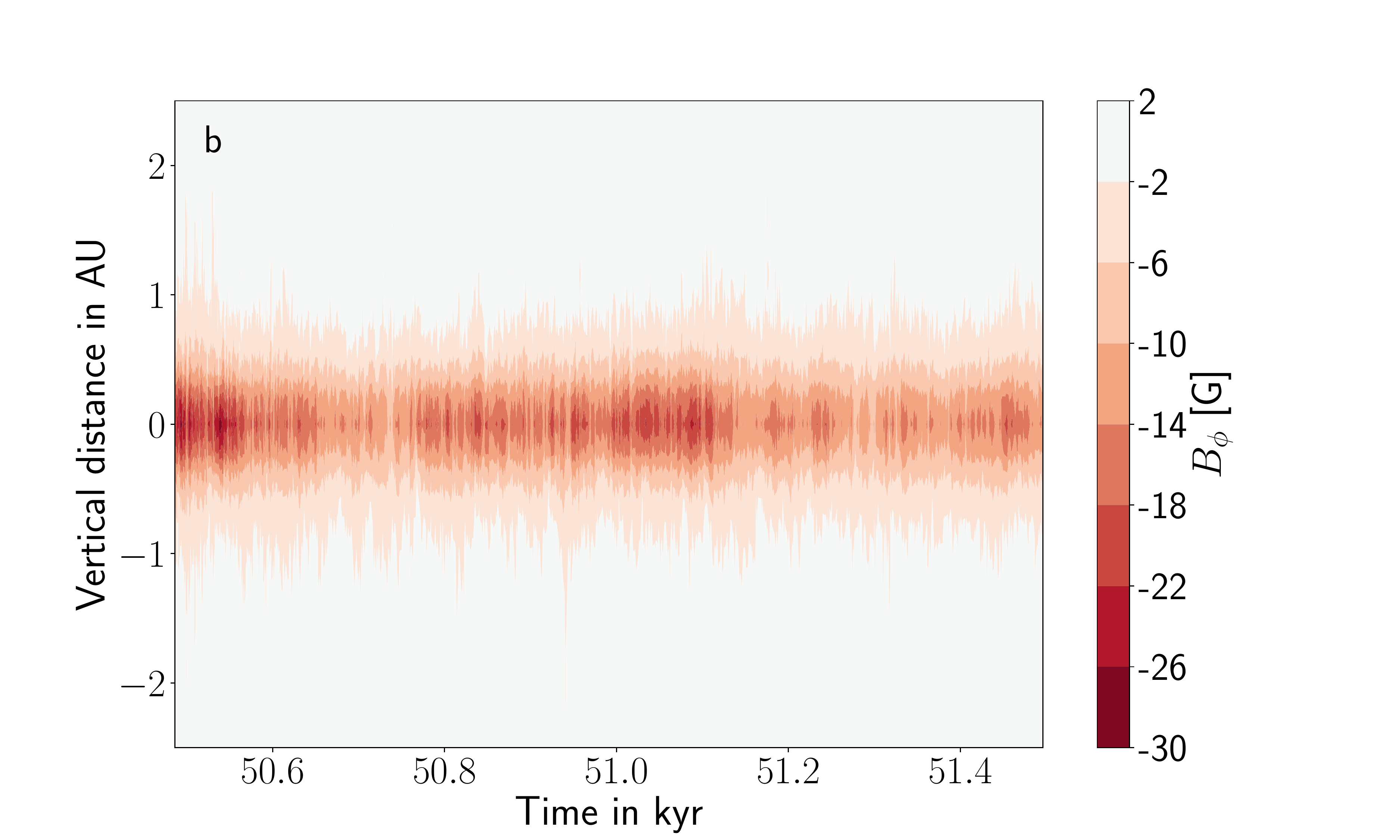}
	\caption{Contour plot of mass-weighted $B_{\phi}$ in $t-z$ plane for the high resolution run. The toroidal magnetic field seems to have focused in the midplane and is predominantly negative with no signs of a sign flip. This is in contrast to weakly magnetised discs that often show sign flips (see \protect\cite{zhustone2017} for one exception).}
	\label{fig:bphihr}
\end{figure}

\Figure{betahr} shows the contour plots of the mass-weighted plasma beta in the high resolution run azimuthally averaged with radial bins of $0.6 \pm 0.1$ AU and vertical bins of $\Delta z \approx 0.2$ AU in each cylindrical shell.  
In the vicinity of the disc midplane, $\beta$ is about $0.01$ -- $0.1$ and decreases to $0.001$ with increasing vertical distance from the midplane up to about $\sim 1$ AU.
This implies that the relative importance of the magnetic field strength increases away from the disc up to a height of about $0.5$ to 1 AU, 
though the magnetisation in absolute terms is strongest in the midplane due the strong toroidal field as shown in \Fig{bphihr}.
An increase of the relative magnetisation with increasing height from the midplane appears to be a generic property of density stratified MHD discs, which might have to do with the exponential drop in thermal pressure and density with height (for isothermal systems).
Even for the weakly magnetised case that is MRI unstable, the relative importance of the $B$-field increases with increasing height, 
\citep[see for example][]{millerstone,nauman2014,suzuki2014,zhustone2017}. 
The key difference in our case is that the discs are more or less strongly magnetised ($\beta \lesssim 1$) everywhere while they are strongly magnetised in the MRI case only above a few scale heights. 

\begin{figure}
	\centering
	\includegraphics[width=0.475\textwidth]{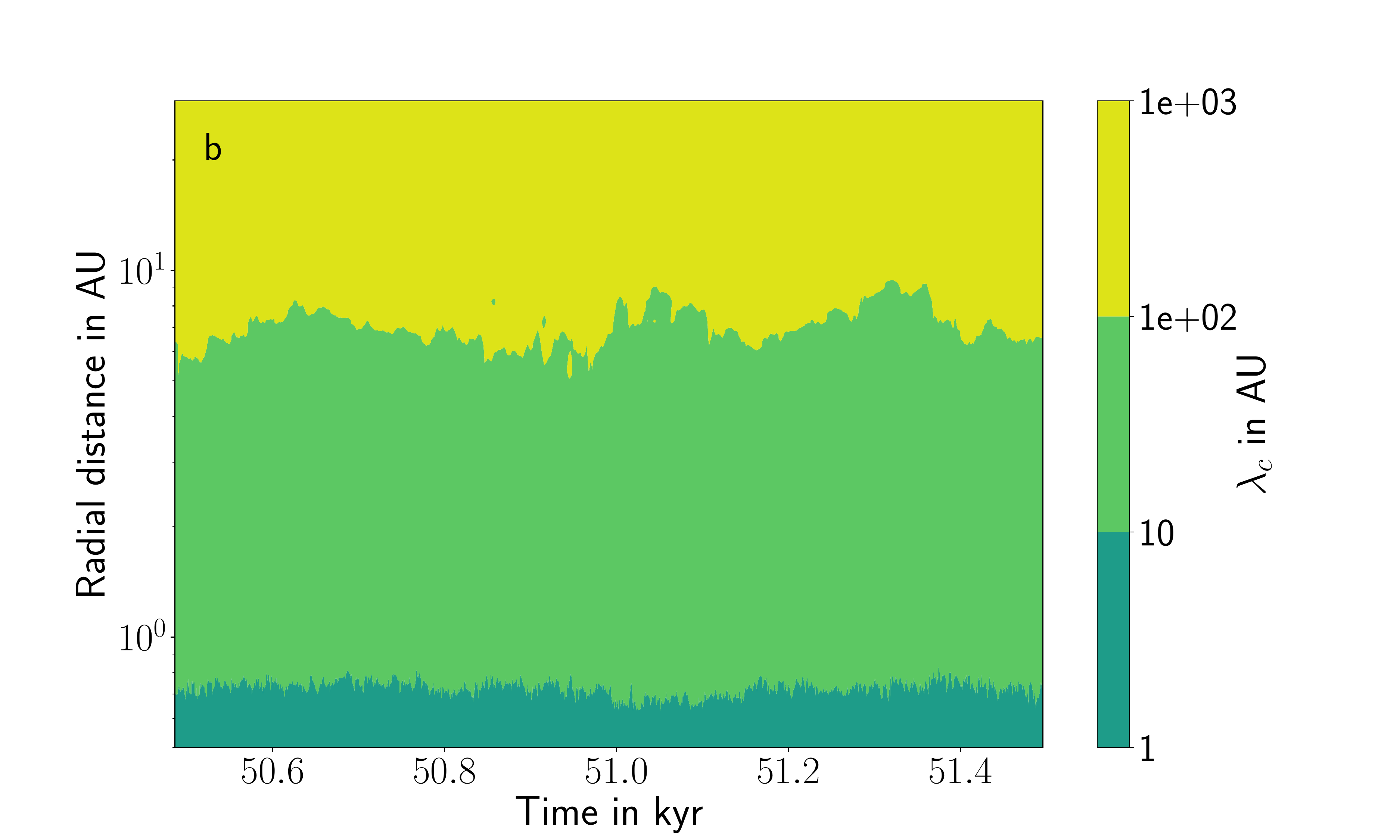}
	\includegraphics[width=0.475\textwidth]{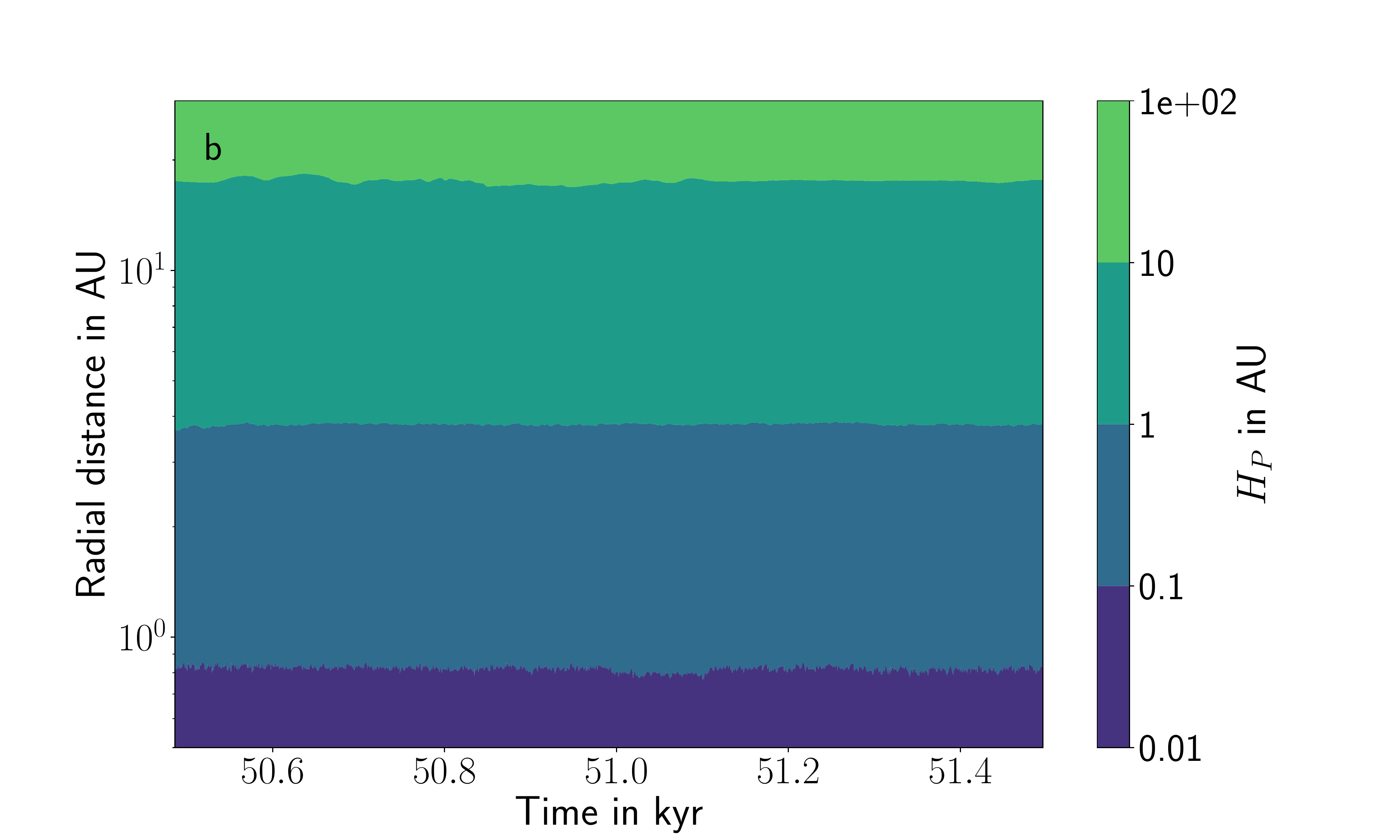}	
	\caption{Plots of the most unstable MRI wavelength $\lambda = 2\pi \sqrt{16/15} |\mathbf{v_A}|/\Omega$ and the thermal pressure scale height $H_P = c_s/\Omega$ for the high resolution run computed as a mass-weighted average in azimuthal bins of $\Delta r \approx 1$ AU and at a vertical extent of $0.2$ AU. We note that the magnetic pressure scale height is considerably larger than the thermal pressure scale height because the Alfv\'en speed in our simulations is considerably larger than the thermal speed.}
	\label{fig:lambda}
\end{figure}

This becomes particularly clear in \Fig{lambda}, where we plot the critical wavelength that corresponds to the most unstable mode of the MRI 
\begin{equation}
\lambda_c = 2\pi \sqrt{16/15} |\mathbf{v}_A|/\Omega,
\label{eq:lambda}
\end{equation}
and the thermal pressure scale height
\begin{equation}
H_P = \frac{c_s}{\Omega}
\label{eq:HB}
\end{equation}
versus radius and time.
In the expressions above, $\mathbf{v}_A$ is the Alfv\'en velocity, $\Omega$ is the orbital frequency and $c_s$ is the sound speed. 
We compute the two quantities in cylindrical shells of size $\Delta r \approx 1, \Delta z \approx \pm 2$ AU. With $\lambda_c$ and $\Delta z$ known, one can compute a quality factor, $Q_{\text{MRI}} = \lambda_c/\Delta z$ \citep{sano2004,hawley2011} to check whether linear modes of MRI due to a purely vertical field are being properly resolved. \Figure{lambda} shows $Q_{\text{MRI}} > 10$ at a few AU from the centre, but we still do not see the MRI develop. This is because $\beta \lesssim 1$ for this disc, and MRI only operates for $\beta \gtrsim 1$. We conclude that the reason MRI does not operate has more to do with the magnetic field strength and not the numerical resolution.

We define the vertical scale height of the magnetic pressure as $H_B = |\mathbf{v}_A|/\Omega$, 
which means that $H_B$ is simply scaled down by a factor of $2\pi \sqrt{16/15}$ compared to $\lambda_c$.
The definitions of the scale heights apply to a thin disc model, while the disc discussed here is relatively thick with 
$\frac{H_P}{r} \sim 0.1$. 
However, the quantities serve reasonably well in obtaining an approximate comparison of the vertical extends of the disc. 
The plots show that especially at small radii, the magnetic pressure scale height is much larger than the thermal pressure scale height in our simulations, a result that can also be seen in the visualisation in \Fig{vapor_vis}. 
We caution the reader that in a more realistic disc simulation, this might not be the case. 
First, the temperatures, and thus the sound speed, would be higher when accounting for the effects of stellar irradiation, especially close to the star.
Second, non-ideal MHD effects -- in particular ohmic dissipation at high densities -- would reduce the magnetic field strength and simultaneously increase the temperatures in the disc.

\begin{figure}
	\centering
	\includegraphics[width=0.475\textwidth]{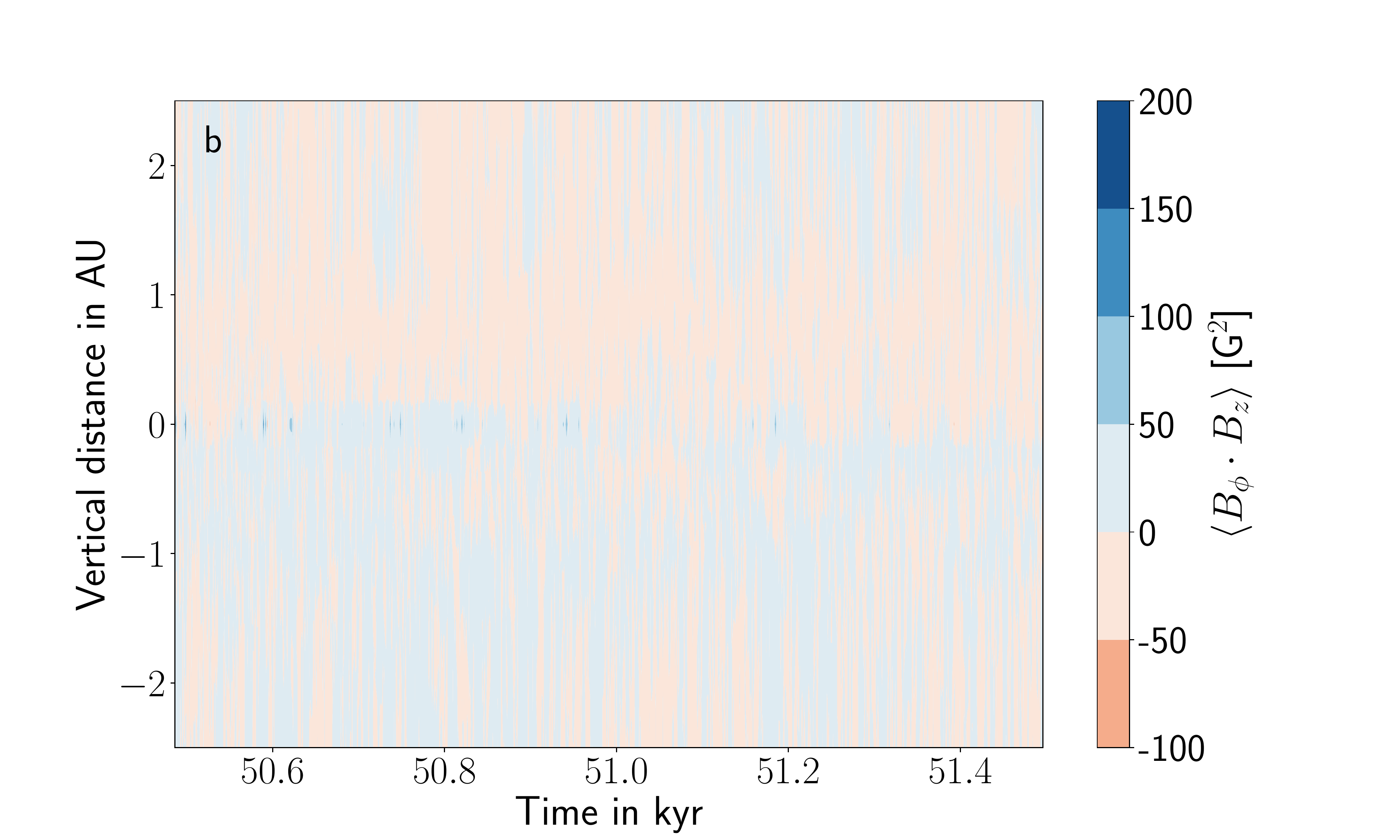}
	\includegraphics[width=0.475\textwidth]{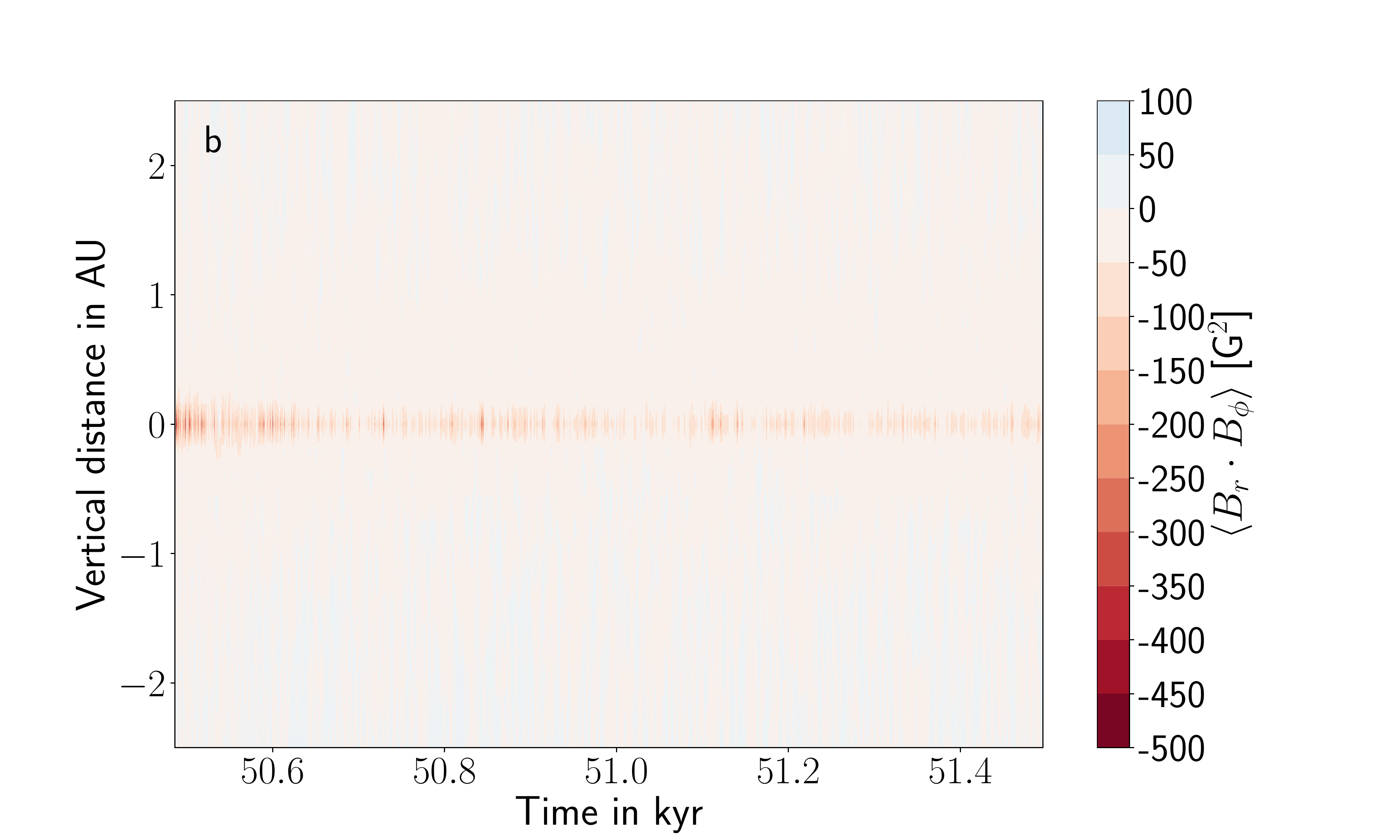}	
	\caption{Contour plot of Maxwell stresses $\lb B_r B_{\phi} \rb$ and $\lb B_{\phi} B_z\rb$ in $t-z$ plane for the high resolution run.}
	\label{fig:maxwellhr}
\end{figure}

As discussed in section 3.4 of K17b, angular momentum in the lower resolution run of the disc is mostly transported via Maxwell stresses predominantly in radial direction. We plot the relevant components of the Maxwell stress ($\lb B_r B_{\phi}\rb$ and $\lb B_z B_{\phi}\rb$) of the high resolution run in \Fig{maxwellhr}. 
$\lb B_r B_{\phi}\rb$ is strongest in the disc region with $- \lb B_r B_{\phi}\rb > 0$ implying angular momentum transport in outward direction in the disc, while $\lb B_r B_{\phi}\rb$ decreases in magnitude and even flips sign for a significant fraction of the vertical box size away from the midplane. 
In the disc region, the vertical-azimuthal Maxwell stress $\lb B_z B_{\phi}\rb$ are smaller in magnitude compared to the radial-azimuthal one. 
We point out that one has to account for a sign flip of the vertical transport depending on whether the angular momentum flux is considered above ($z>0$) or below the disc ($z<0$). 
Therefore, $\lb B_z B_{\phi}\rb < 0$ for $z>0$ corresponds to outward transport because the Maxwell stress through a surface above the midplane 
is $-\lb B_z B_{\phi}\rb > 0$, whereas $\lb B_z B_{\phi}\rb < 0$ for $z<0$ corresponds to inward transport 
through a surface below the midplane.

As seen in the upper panel of \Fig{maxwellhr}, the overall quantity $\lb B_z B_{\phi}\rb$ is predominantly negative for $z>0$ and predominantly positive for $z<0$. 
Accounting for the sign flip and the minus sign in the definition of the Maxwell stress, this means that angular momentum is transported outwards via vertical magnetic fields.
This result is in good agreement with the previous results found in K17b regarding angular momentum transport.
Therefore, these results show that outward angular momentum transport is controlled by Maxwell stresses, 
and thus induced by the globally inherited magnetic field. For the short time that we evolve the highly resolved disc, we do not observe sign flips in the azimuthal magnetic field in the $t-z$ plane shown in \Fig{bphihr}. However, notice that the same disc does show a sign flip on longer time scales as evident in \Fig{bphi} (see sign flip at about $t=10$ kyr for sink b). This is characteristic of strongly magnetised discs as reported in \cite{salvesen2016}.

\subsection{Comparison with local and global disc simulations}
One might ask what is the advantage of using a star formation simulation to zoom-in on accretion discs over doing local and global disc simulations? The biggest advantages are that we avoid artificial initial and outer boundary conditions -- in commonly applied setups of global disc simulations, the initial conditions describe a disc in equilibrium that is then perturbed, and the radial/vertical boundaries are typically assumed to be inflow/outflow \citep{parkin2014,suzuki2014,zhustone2017}.
In the K17 simulations, the outer boundaries and the initial conditions are determined self-consistently as a result of stellar collapse with the caveat that these current simulations do not include potentially important effects such as non-ideal MHD effects or radiation transport. 
However, as in global disc simulations, the inner radial boundary (sink particle) remains an idealisation considering the
real properties of a star though typically only the dynamics within a few cell lengths from the sink are spuriously affected by the sink (see K17b).  

Most previous works on strongly magnetised discs \citep{johansenlevin2008,salvesen2016,gaburov2012,fragile2017} have focused on the effect of field topology and orientation, for example 
a purely toroidal field versus a purely vertical field. The magnetic fields present in our discs have both a vertical and a toroidal part. 
We also note that our simulations are evolved for much longer time scales than local and global disc simulations, which typically are only evolved up to a few hundred local orbits.

\section{Conclusions}

Our hyper-global simulations demonstrate the importance of large-scale magnetic fields in young circumstellar discs,
and thus, the necessity of implying their effects in realistic disc models.  
In particular, our high resolution run suggests that the hyper-global magnetic fields 
cause a high enough magnetic field strength in the disc to prevent the onset of the MRI.
Given that we neither account for non-ideal MHD effects nor radiative transfer in our models, 
we most likely overestimate the (relative) importance of the magnetic field strength.
Nevertheless, the results show that large-scale magnetic fields influence the properties of young discs,
and they suggest that young circumstellar discs may be more magnetised 
than assumed in common disc models that assume later stages of disc evolution.
It will be interesting to see whether a strong magnetic field in young discs diffuses away during the evolution of the disc,
or whether non-ideal MHD effects (in particular ambipolar diffusion) 
prevent the formation of strongly magnetised discs altogether.
Observationally, the former would mean that an unknown fraction of discs would be strongly magnetised, 
and the magnetic field strength could thus be used for tracing the age of discs.
We think that our hyper-global approach is complementary to the more standard procedure of focusing on discs exclusively,
and particularly useful to understand magnetic fields in accretion discs.

\section*{Acknowledgments}
We thank Troels Haugb\o lle and \AA ke Nordlund for their code development that allowed MK to carry out the simulations. 
Moreover, we thank Troels Haugb\o lle for comments on an earlier draft.
MK's research at the Centre for Star and Planet Formation is funded by the Danish National Research Foundation (DNRF97). 
MK acknowledges PRACE for awarding us access to the computing resource CURIE based in France at CEA for carrying out part of the simulations. Archival storage and computing nodes at the University of Copenhagen HPC center, funded with a research grant (VKR023406) from Villum Fonden, were used for carrying out part of the simulations and the post-processing.
FN acknowledges support from the European Research Council under the European Union Seventh Framework Programme (FP/2007-2013) under ERC grant agreement 306614.
Finally, we acknowledge the developers of the 
python-based analysis tool yt (http://yt-project.org/) \citep{yt-reference}. 


\label{lastpage}

\end{document}